\documentclass[11pt,prd,a4paper,preprintnumbers,amsmath,amssymb,nofootinbib]{article}
\pdfoutput=1
\usepackage{feynmp-auto,expdlist}
\usepackage{amsmath,amsfonts,amssymb}
\usepackage{graphicx}
\usepackage{enumerate}
\usepackage{hyperref}
\usepackage{latexsym}
\usepackage{hepnicenames}
\usepackage{enumerate}
\usepackage{soul}
\usepackage[normalem]{ulem}
\usepackage{wasysym}
\usepackage{makecell}
\usepackage{bbm}
\usepackage{physics}

\font\tenrsfs=rsfs10 at 12pt
\font\sevenrsfs=rsfs7
\font\fiversfs=rsfs5
\newfam\rsfsfam
\textfont\rsfsfam=\tenrsfs
\scriptfont\rsfsfam=\sevenrsfs
\scriptscriptfont\rsfsfam=\fiversfs

\numberwithin{equation}{section}

\usepackage{mathrsfs}
\usepackage{braket}
\usepackage{titling}
\usepackage{amsmath}
\usepackage{slashed}
\usepackage{amssymb}
\usepackage{epsfig}
\usepackage{graphicx}
\usepackage{color}
\usepackage{rotating}
\usepackage{hyperref}
\usepackage[margin=1.in]{geometry}
\usepackage[table,xcdraw,dvipsnames]{xcolor}
\usepackage[numbers,sort&compress]{natbib} 
\usepackage{colortbl}
\usepackage{pdflscape}
\usepackage{color}
\usepackage{mathtools}
\usepackage{colortbl}
\usepackage{comment}

\renewcommand{\H}{{\cal H}}

\newcommand{\U}{{\rm U}}

\newcommand{\V}{{\cal V}}

\definecolor{nicered}{rgb}{0.7,0.1,0.1}
\definecolor{nicegreen}{rgb}{0.1,0.5,0.1}
\definecolor{red}{rgb}{1.0, 0, 0}
\definecolor{niceblue}{rgb}{0,0,0.8}
\allowdisplaybreaks

\definecolor{blus}{cmyk}{1,1,0,0.6}
\definecolor{verde}{cmyk}{0.92,0,0.59,0.25}
\definecolor{rossos}{cmyk}{0,1,1,0.55}

\hypersetup{colorlinks,bookmarksopen,bookmarksnumbered,linkcolor=blus,pdfstartview=FitH,urlcolor=rossos,citecolor=verde}

\def\eq#1{{Eq.~(\ref{#1})}}
\def\eqs#1#2{{Eqs.~(\ref{#1})--(\ref{#2})}}
\def\fig#1{{Fig.~\ref{#1}}}

\def\sect#1{{Section~\ref{#1}}}

\def\app#1{{Appendix~\ref{#1}}}

\def\vev#1{\left\langle #1\right\rangle}

\renewcommand{\bar}{\overline}

\newcommand{\beq}{\begin{equation}}
\newcommand{\eeq}{\end{equation}}
\newcommand{\bea}{\begin{eqnarray}}
\newcommand{\eea}{\end{eqnarray}}

\renewcommand{\[}{\left[}
\renewcommand{\]}{\right]}
\renewcommand{\(}{\left(}
\renewcommand{\)}{\right)}

\def\be{\begin{equation}}
\def\ee{\end{equation}}

\begin{document}

\begin{center}  
{\LARGE
\bf\color{blus} 
On the 
oscillating 
electric dipole moment \\ \vspace{0.2cm}
induced by axion-fermion couplings} \\
\vspace{0.8cm}

{\bf Luca Di Luzio, Hector Gisbert, Philip S{\o}rensen}\\[7mm]

{\it Istituto Nazionale di Fisica Nucleare (INFN), Sezione di Padova, \\
Via F. Marzolo 8, 35131 Padova, Italy}\\[1mm]
{\it Dipartimento di Fisica e Astronomia `G.~Galilei', Universit\`a di Padova,
 \\ Via F. Marzolo 8, 35131 Padova, Italy
}\\[1mm]

\vspace{0.3cm}
\begin{quote}

It has been recently claimed that the axion coupling to fermions is responsible for an oscillating electric dipole moment (EDM) in the background of axion dark matter. In this work, we re-examine the derivation of this effect. Contrary to previous studies, we point out the physical relevance of an axion boundary term, which is crucial in restoring the axion shift symmetry and drastically affects the EDM phenomenology. To describe the latter, we introduce the notion of a time-averaged effective axion EDM, which encodes the boundary term and whose magnitude depends on the oscillation regime. For slow oscillations, the boundary term washes out the standard oscillating EDM, resulting in an exact cancellation in the static limit. Conversely, during fast oscillations, the boundary term amplifies the effective EDM relatively to the standard EDM contribution. This observable is especially interesting in the case of the electron EDM. 
For an $\mathcal{O}(1)$ axion-electron coupling, the overall size of the effective EDM in the regime of intermediate or fast oscillations is comparable to the present static EDM limit.

\end{quote}

\thispagestyle{empty}

\end{center}

\bigskip
\tableofcontents

\clearpage

\section{Introduction}
\label{sec:intro}

The QCD
axion 
\cite{Peccei:1977hh,Peccei:1977ur,Weinberg:1977ma,Wilczek:1977pj}
represents a most compelling paradigm 
for the physics beyond the Standard Model (SM), 
by simultaneously providing an elegant 
solution to the strong CP problem 
and an excellent dark matter candidate \cite{Preskill:1982cy,Abbott:1982af,Dine:1982ah}.
Current axion searches have started to probe the region of parameter space predicted by the QCD axion 
(see Ref.~\cite{DiLuzio:2020wdo} for a comprehensive overview of axion models) while 
next-generation experiments 
will explore 
regions of parameters space deemed to be unreachable until a decade ago \cite{Irastorza:2018dyq,Sikivie:2020zpn}.

Among novel and ingenious axion detection strategies, 
it was proposed in Refs.~\cite{Graham:2013gfa,Budker:2013hfa,JacksonKimball:2017elr} to employ 
nuclear magnetic resonance techniques 
to look 
for an oscillating neutron 
electric dipole moment (EDM) 
induced by axion dark matter. 
Other experimental approaches 
have been suggested in order to 
search for oscillating EDMs 
of atoms, molecules and nuclei \cite{Stadnik:2013raa,Abel:2017rtm,Flambaum:2019ejc}, 
as well as that of protons/deuterons \cite{Chang:2017ruk,Pretz:2019ham,Graham:2020kai,Kim:2021pld}
and electrons \cite{Suleiman:2021whz}
employing storage ring facilities. 
As a theoretical setup, most of these 
works focussed on the oscillating nucleon EDM 
induced by the defining property of the QCD axion, namely an axion coupling to the $G\tilde G$ operator. 

On the other hand, it has been recently 
pointed out 
by 
Smith \cite{Smith:2023htu} 
(see also Refs.~\cite{Alexander:2017zoh,Wang:2021dfj} 
for earlier similar claims)
that 
the model-dependent axion-fermion coupling, 
here denoted as $g = C \, m /f_a$ 
(where $C$ is an adimensional 
coupling, $m$ the fermion mass and $f_a$ the axion decay constant)
leads 
to an axion-dependent EDM of the type 
\beq 
\label{eq:aedm}
 d(a) = \frac{e \,g }{2\, m^2}\,a = 
\frac{C \,e }{2\, m}\,\frac{a}{f_a}
\,, 
\eeq
defined in terms of the non-relativistic 
(NR) Hamiltonian,  
$\H_{\rm EDM} \supset - d(a)\,\vec{\sigma}\cdot\vec{E}$.

\eq{eq:aedm} would have striking experimental consequences,  
since in the background of axion dark matter, 
with the amplitude of the oscillation fixed in terms of the local dark matter relic density, 
one predicts in the case of electrons 
an oscillating 
EDM, $d_e (a) \simeq  7 \cdot 10^{-30}\, C\, \cos(m_a t)$ e cm \cite{Smith:2023htu},
that for $C \sim 1$ is 
comparable 
the current limit on the static EDM, 
$|d_e| \lesssim 1.1 \cdot 10^{-29}$ e cm
\cite{ACME:2018yjb}.

This paper aims to 
re-examine 
the derivation of the axion-induced EDM in 
\eq{eq:aedm}, 
given its potential 
impact
for EDM searches. 
While the NR expansion of the 
axion Dirac equation in Ref.~\cite{Smith:2023htu} 
was obtained by means of the Foldy-Wouthuysen approach \cite{Foldy:1949wa}, 
which consists of a block-diagonalization of the Dirac equation, 
we here employ the more direct 
Pauli elimination method \cite{Pauli:1927qhd} in order to provide an 
independent derivation. 
These two techniques are known to yield 
equivalent results
(see e.g.~\cite{deVries:1968ksb}), 
although the Pauli elimination method requires careful treatment of spinor normalization.

Throughout our derivation, we pay a special attention 
to verifying that the expected properties 
of the axion theory, 
such as the equivalence of the axion formulation  
in the derivative and exponential bases, as 
well as the axion shift symmetry, 
are properly satisfied. 
While we were able to reproduce most of the results 
of Ref.~\cite{Smith:2023htu} (as well as \cite{Alexander:2017zoh,Wang:2021dfj}), 
we disagree 
on the physical interpretation of the 
axion-fermion coupling 
in terms of a standard axion EDM, 
as given by \eq{eq:aedm}. 
The crucial point consists in the identification 
of a previously overlooked axion boundary term 
which is 
needed 
in order 
to restore the axion shift symmetry, 
irrespective of the chosen axion basis.
Without this term, the axion shift symmetry  
would be
explicitly 
broken by \eq{eq:aedm}. 
Remarkably, the inclusion of this boundary term in time-dependent perturbation 
theory drastically affects the phenomenology 
of the standard axion EDM. 

Nonetheless, in the 
presence of a constant electric field,  
it is still possible to define a 
 time-averaged effective
EDM, which is made of the standard EDM of \eq{eq:aedm} 
plus a contribution arising from the axion boundary 
term. Depending on the oscillation regime, 
different behaviors emerge. 
For slow oscillations the 
axion boundary term washes out 
the standard oscillating EDM, resulting in an exact cancellation in the static limit. In the 
regime of fast oscillations, 
where the standard EDM contribution is suppressed,
the axion boundary term 
instead maintains a constant amplitude of the effective EDM. 
For intermediate or fast oscillations,
the effective EDM is similar in size 
to the standard one 
in \eq{eq:aedm}.  

The paper is structured as follows. We start in 
\sect{sec:axionbases} by presenting the axion Lagrangian in the exponential and derivative bases. 
In \sect{sec:NRexp} we provide the NR expansion 
of the axion Dirac equation in the derivative basis, 
while we refer to \app{app:equivalence} 
for the calculation in the exponential basis. 
In \sect{sec:unitary} we discuss the formal 
equivalence of 
the axion Hamiltonian as derived in 
the exponential and the derivative bases.
We discuss this by means of a unitary transformation 
that is derived from the NR limit of the 
field redefinition connecting the two bases 
at the Lagrangian level. Next, 
in \sect{sec:timedepPT}, 
we consider the consequences of the axion Hamiltonian 
in time-dependent perturbation theory, we argue that the axion boundary term is physically relevant, and finally, we introduce the concept of a 
time-averaged effective
axion EDM.
We conclude in \sect{sec:concl}, 
summarizing the main findings of our study.

\section{Axion Lagrangian: 
exponential vs.~derivative basis} 
\label{sec:axionbases}

Let us consider 
a toy model with 
a massless 
axion interacting with a Dirac fermion field 
charged under an abelian gauge group (e.g.~an electron).  
In specific ultraviolet completions, 
the axion can be represented as the 
orbital Goldstone mode of a complex scalar field, 
$\phi \supset f_a e^{i a / f_a}$, 
whose radial mode has been integrated out. 
This defines the axion effective Lagrangian in the \textit{exponential basis}
\begin{align} 
\label{eq:Laxionnonder}
\mathcal{L}_E &\,=\, \frac{1}{2} \,(\partial_\mu a)^2\, 
+\, \bar \psi_E\, i\, \slashed{\partial}  \,\psi_E\, +\, e\, A_\mu\, \bar \psi_E\,\gamma^\mu \,\psi_E\, -\, (m\, \bar \psi_{EL}\, \psi_{ER}\, 
e^{i \frac{g}{m} a} 
\,+\, \text{h.c.}) \, ,     
\end{align}
where we have introduced the coupling $g$, 
defined via $g/m = 1/f_a$.\footnote{In 
realistic QCD axion models 
with axion couplings to SM fermions,
such as the DFSZ model \cite{Zhitnitsky:1980tq,Dine:1981rt},
the axion spans over different Higgs 
multiplets and $g/m$ is related to the 
axion decay constant via $g/m = C/f_a$, 
where $C$ is an $\mathcal{O}(1)$ 
parameter proportional to the PQ charge 
of the associated Higgs doublet (see e.g.~Sect.~2.7.2 in \cite{DiLuzio:2020wdo}).} 
The subscript $_E$ is a reminder that these fields are defined in the exponential basis and $_L$ and $_R$ indicate left and right chiral components, respectively. The $\U(1)_{\rm PQ}$ 
symmetry is implemented as 
\beq 
\label{eq:PQsymmetryEXP}
\U(1)^{E}_{\rm PQ}: \qquad
a \to a + \alpha \, m / g \, , \quad 
\psi_E \to e^{-i \frac{\alpha}{2} \gamma_5} \psi_E \, ,  
\eeq
with $\alpha$ the global transformation parameter.
The last term in the bracket of \eq{eq:Laxionnonder} can be also expressed as 
\begin{align}
   \begin{aligned}
-\, (m \,\bar \psi_{EL}\, \psi_{ER} e^{i\frac{g}{m} a}\, +\, \text{h.c.}) 
&\,=\, -\, m\, \cos \(\frac{g}{m} a\)\, \bar \psi_E \,\psi_E\, -\,  i\, m\, \sin \(\frac{g}{m} a\)\, \bar \psi_E \, \gamma_5 \,\psi_E \label{eq:introduction of trig} \\
&\,\simeq\,  -\, m\, \bar \psi_E\, \psi_E\, -\,  g\, a\, \bar \psi_E\, i\, \gamma_5\, \psi_E \, , 
\end{aligned} 
\end{align}
where in the last step we have expanded at the first non-trivial order in the axion field, 
    which defines the so-called \textit{linear basis}. 
Note that the linear approximation might lead to incorrect results if more than one axion is involved in a given process (for an explicit example, 
cf.~the discussion around \eq{eq:Hlinbasis}). 

\eq{eq:Laxionnonder} can be brought into the \textit{derivative basis} via an axion-dependent field redefinition:
\beq 
\label{eq:fieldred}
\psi_E \to e^{i \gamma_5 \frac{g}{2 m} a} \psi_E 
\equiv \psi_D \, ,
\eeq
so that $\mathcal{L}_E$ is mapped into the equivalent Lagrangian 
\begin{align}
\label{eq:Laderivbasis}
\mathcal{L}_E\, \to\, \mathcal{L}_D &\,=\, \frac{1}{2}\, (\partial_\mu a)^2\, 
+\, \bar \psi_D\, (i\, \slashed{\partial} \,-\, m)\, \psi_D\, +\, e\, A_\mu\, \bar \psi_D\,\gamma^\mu\, \psi_D \nonumber \\
&\,+\, \frac{g}{2m}\, (\partial_\mu a) \,\bar \psi_D\, \gamma^\mu\, \gamma_5\, \psi_D \,
+\, \frac{e^2}{16 \pi^2}\, \frac{g}{2m}\, a\, F^{\mu\nu}\, \tilde F_{\mu\nu} \, ,  
\end{align}
with the last term 
arising from the non-invariance of the path-integral 
measure \cite{Fujikawa:1979ay}
under the anomalous field redefinition in 
\eq{eq:fieldred}. 
In the derivative basis  
the original 
$\U(1)_{\rm PQ}$ 
symmetry defined in \eq{eq:PQsymmetryEXP}
is implemented as 
\beq 
\label{eq:PQsymmetryDER}
\U(1)^{D}_{\rm PQ}: \qquad
a \to a + \alpha \, m /g \, , \quad 
\psi_D \to \psi_D \, ,  
\eeq
with the $\psi_D$ field that is left invariant. 

In the following, we will 
perform the NR limit of the 
axion Dirac equation 
in the derivative basis,   
and compare this to the calculation 
in the exponential basis. 
After showing the equivalence of the 
two formulations, we will finally discuss 
the physical consequences 
of the derived axion Hamiltonian 
in time-dependent perturbation theory.

\section{Non-relativistic limit of the axion Dirac equation}

\label{sec:NRexp}

The axion Dirac equation 
stemming from 
\eq{eq:Laderivbasis} 
in the derivative basis 
reads
\beq 
\label{eq:Diraceqder}
\left(i\, \gamma^\mu\,\partial_\mu\, -\, m\, +\, e\,\gamma^\mu\, A_\mu\, +\,\frac{g}{2m}\,(\partial_\mu a)\,\gamma^\mu\, \gamma_5 \right)\,\psi\, =\, 0 \, ,
\eeq 
where we dropped for simplicity the subscript $_D$ on the spinor field.
To perform the NR limit of the axion Dirac equation, it is convenient to employ the Dirac representation for the gamma matrices
\beq
\gamma^0 = 
\begin{pmatrix}
\mathbf{1} & 0 \\
0 & -\mathbf{1}
\end{pmatrix} \, , \quad 
\gamma^i = 
\begin{pmatrix}
0 & \sigma^i \\
-\sigma^i & 0
\end{pmatrix} \, , \quad 
\gamma^5 = 
\begin{pmatrix}
0 & \mathbf{1} \\
\mathbf{1} & 0 
\end{pmatrix} \, ,
\label{eq:gamma matrices}
\eeq
with $\sigma^{i=1,2,3}$ denoting the Pauli matrices,  
and 
adopt the subsequent parametrization for the Dirac spinor 
\begin{align}\label{eq:usualansatz}
\psi\,(\vec{x},t) = \text{exp}\{-i\,m\,t\}\, 
\begin{pmatrix}
\chi\\
\Phi
\end{pmatrix}~. 
\end{align}
The mass term in \eq{eq:usualansatz}
provides the dominant time evolution in the NR limit, 
while the bi-spinors $\chi=\chi\,(\vec x, t)$ and $\Phi=\Phi\,(\vec x, t)$
are slowly varying functions 
which exhibit  
a small energy dependence suppressed by $1/m$. 
In terms of two-component spinors
the axion Dirac equation~\eqref{eq:Diraceqder} 
can be written as
\begin{align}\label{eq:Diracmatrix}
\begin{pmatrix}
\hat{A} & \hat{B} \\
\hat{C} & \hat{D} 
\end{pmatrix}
\begin{pmatrix}
\chi \\
\Phi
\end{pmatrix}
= 0~,
\end{align}
with 
\begin{align}\label{eq:ABCD}
    \hat{A}\,&=\,\hat{\epsilon}\,+\,\frac{g}{2\,m}\,\vec{\sigma}\cdot\vec{\nabla}a\,+\,e\,A_0~,\nonumber\\
    \hat{B}\,&=\,-\,\vec{\sigma}\cdot\hat{\vec{P}}\,+\,\frac{g}{2\,m}\,\dot{a}~,\nonumber\\
    \hat{C}\,&=\,\vec{\sigma}\cdot\hat{\vec{P}}\,-\,\frac{g}{2\,m}\,\dot{a}~,\nonumber\\
    \hat{D}\,&=\,-\left(2\,m\,+\,\hat{\epsilon}\,+\,\frac{g}{2\,m}\,\vec{\sigma}\cdot\vec{\nabla}a\,+\,e\,A_0\right)~,
\end{align}
where operators are denoted with a hat,  
$\hat{\vec{P}}=\hat{\vec{p}}+e\,\vec{A}$ with $\hat{\vec{p}}=-\,i\,\vec{\nabla}$ and  $\hat{\epsilon}=\,i\,\partial_t$.

To perform the NR expansion 
we employ the Pauli elimination method~\cite{Pauli1990-jc}, which 
consists in 
substituting the lower equation $\Phi=-\hat{D}^{-1}[\hat{C}[\chi]]$ from~\eq{eq:Diracmatrix} into the upper one,\footnote{Since $\Phi 
\simeq 
(\vec{\sigma} \cdot \hat{\vec{p}} + \dots) / (2m)\,\chi \ll \chi$ in the NR limit,  
$\chi$ and $\Phi$ go under the name of large and small 
component spinors, respectively.} 
thus obtaining:
\begin{align}\label{eq:DiracAxion}
\hat{A}[\chi] - \hat{B}[\hat{D}^{-1}[\hat{C}[\chi]]]=0~,
\end{align}
where to keep track of the operatorial nature of $\hat{\vec{p}}$ and $\hat{\epsilon}$
we have introduced the notation $X[...]$, meaning that ``$X$'' acts on ``$...$''.

We next expand the inverse of $\hat{D}=-(2\,m\,+\,\hat{\epsilon} +\,\frac{g}{2\,m}\,\vec{\sigma}\cdot\vec{\nabla}a\, + e A_0)
=-\,2\, m\,(1+\,(\hat{\epsilon} + e A_0)/(2\,m)+\mathcal{O}(1/m^2))$. Therefore, 
when applied to ``$x$'', 
we obtain:
\begin{align}\label{eq:Dinverse}
   \hat{D}^{-1}[x]\,=\,-\, \frac{1}{2\,m} &\left( 
  x - \frac{\hat{\epsilon}[x] + e\, A_0[x]}{2 m}\right)\,+\,\mathcal{O}(1/m^3)~.
\end{align}
This expression provides a solution for the inverse of $\hat{D}$ at second order, with further terms in the expansion accounting for $\mathcal{O}(1/m^3)$ higher-order corrections. Including all the previous ingredients in Eq.~\eqref{eq:DiracAxion}, 
we match the latter onto the form of a 
time-independent 
Schr\"odinger equation, given by
\beq 
\hat{\H}[\chi]\,=\,\left(\hat{\H}^{(0)}\,+\,\hat{\H}^{(1)}\,+\,...\right)[\chi]\,=\,\hat{\epsilon}[\chi] \, ,  \label{eq: Schrodinger}
\eeq
where the superscript $(i)$ denotes terms at $\mathcal{O}(1/m^i)$. 

In the following, it will be useful to split 
$\hat{\H}^{(i)}\,=\,\hat{\H}^{(i)}_a\,+\,\hat{\H}^{(i)}_\slashed{a}$, respectively into a piece with 
and without the axion. 
At $\mathcal{O}(1/m)$ we obtain the axion Hamiltonian,
\begin{align}\label{eq:nonormalizedaxionEXPbasis}
    \hat{\H}^{(1)}_a\,=\,-\,\frac{i\,g\,[\vec{\sigma}\cdot\hat{\vec{P}},a]}{2\,m}\,
    =\,-\frac{g}{2\,m}\vec{\sigma}\cdot\vec{\nabla}a~, 
\end{align}
which corresponds to the so-called axion-wind term 
(see e.g.~\cite{Graham:2013gfa}). 
Note that 
in the last step 
of \eq{eq:nonormalizedaxionEXPbasis}
we exploited the relation
\begin{align}\label{eq:algebra}
    [\vec{\sigma}\cdot\hat{\vec{P}},a]\, f\,&=\,\vec{\sigma}\cdot\hat{\vec{P}}\,[a\,f]-a\,\vec{\sigma}\cdot\hat{\vec{P}}\,[f]\, =\,\vec{\sigma}\cdot\hat{\vec{p}}\,[a\,f]-a\,\vec{\sigma}\cdot\hat{\vec{p}}\,[f] \nonumber \\
    &=\,(\vec{\sigma}\cdot\hat{\vec{p}}\,a)\,f\,=\,-\,i\,(\vec{\sigma}\cdot\vec{\nabla}\,a)\,f~,
\end{align}
with $f$ an auxiliary function and $[a,b] \,c\equiv a[b[c]] - b[a[c]]$. 

The previous calculation can be extended at $\mathcal{O}(1/m^2)$. 
From Eq.~\eqref{eq:DiracAxion} we obtain  $\hat{\mathcal{H}}[\chi]=\hat{\epsilon}[\chi]$, with
\begin{equation}\label{eq:fuA}
\boxed{
\hat{\H}^{(2)}_{a} =-\,\frac{g\,\lbrace\vec{\sigma}\cdot\hat{\vec{P}},\dot{a}\rbrace}{4\,m^2} } = - \,\frac{g \,\dot a\, \vec{\sigma} \cdot \hat{\vec{P}}}{2\,m^2}\, 
+i\, \frac{g\, \vec{\sigma} \cdot{\vec{\nabla}}\,\dot a}{4\, m^2}\, 
\, , 
\end{equation}
where in the last step we used 
a relation analogous to \eq{eq:algebra}
and 
$\lbrace a,b\rbrace \,c\equiv a[b[c]] + b[a[c]]$.  
Note that the $\dot a\, \vec{\sigma} \cdot \hat{\vec{p}}$ term stemming from 
\eq{eq:fuA}
leads to the so-called axioelectric effect, 
discussed e.g.~in Ref.~\cite{Pospelov:2008jk}. 

On the other hand, the calculation in the 
exponential basis 
leads to a different Hamiltonian 
starting at $\mathcal{O}(1/m^2)$, 
see \app{app:equivalence} 
for details.\footnote{A non-trivial issue 
that one has to deal with in the exponential basis 
is the fact that, by simply employing the 
Pauli elimination method, the Hamiltonian 
at $\mathcal{O}(1/m^2)$ is not Hermitian. 
An analogous problem arises 
for the correct identification of the Darwin term 
in the NR Hamiltonian 
describing 
the hydrogen fine structure 
(see e.g.~\cite{shankar1994principles}) 
and its solution requires 
an extra normalization 
of the Schr\"odinger spinor. 
In Appendix \ref{app:equivalence}, we provide a comprehensive discussion
 of this normalization issue, both for the exponential and derivative bases. For the latter basis, 
 the problem 
 arises only at $\mathcal{O}(1/m^3)$ 
 and hence it does not 
 show up in the present calculation.} 
Denoting the axion Hamiltonian 
in the 
exponential basis as $\hat\H_{Ea}$, we obtain
\begin{equation}
\label{eq:Hexp}  
\boxed{
   \hat\H^{(2)}_{Ea} =-\,\frac{g\,\lbrace\vec{\sigma}\cdot\hat{\vec{P}},\dot{a}\rbrace}{4\,m^2} 
   - \frac{eg}{4m^2} a \vec{\sigma}\cdot\vec{E} +\frac{g}{4m^2}\dot{a}\vec{\sigma}\cdot\hat{\vec{P}}\, ,} 
\end{equation}
which differs from \eq{eq:fuA} 
by the last two terms 
and, in particular, 
it features  
an axion EDM term 
proportional to the electric field, 
$\vec{E} = - \,\partial_t \, \vec{A} - \,\vec{\nabla} A_0$.

\section{Equivalence of  
bases from unitary transformations}
\label{sec:unitary}

Before addressing the 
physical 
consequences of the
axion Hamiltonian at $\mathcal{O}(1/m^2)$, 
we wish to discuss the 
equivalence of the 
two formulations 
in the derivative and exponential 
bases. 
At the formal level, it is possible to show that 
the Hamiltonians 
in \eqs{eq:fuA}{eq:Hexp} 
are connected by a unitary transformation, that is nothing but the 
NR limit of the field redefinition in \eq{eq:fieldred}, 
allowing us to go from the exponential to the derivative basis. 

To this end, let us first rewrite \eq{eq:fieldred}
in terms of bi-spinors as 
\begin{align}
\label{eq:EtoD}
\begin{pmatrix}
\chi_E \\
\Phi_E
\end{pmatrix}\to\text{exp}\left\lbrace i\,\gamma_5\,\frac{g\,a}{2\,m}\right\rbrace\,
\begin{pmatrix}
\chi_E \\
\Phi_E
\end{pmatrix}
\,
\equiv\,
\begin{pmatrix}
\chi_D \\
\Phi_D\end{pmatrix}~, 
\end{align}
where we have explicitly labeled 
the states 
in the exponential ($E$) and derivative ($D$) bases.
Expanding the above expression in powers of $1/m$, 
we obtain:
\begin{align}
\label{eq:uniteqexp}
\begin{pmatrix}
\chi_E \\
\Phi_E
\end{pmatrix}\,=\,\text{exp}\left\lbrace -\,i\,\gamma_5\,\frac{g\,a}{2\,m}\right\rbrace\,\begin{pmatrix}
\chi_D \\
\Phi_D\end{pmatrix}\,=\,\begin{pmatrix}
1 & -\,i\,\frac{g}{2\,m}\,a\\
-\,i\,\frac{g}{2\,m}\,a & 1
\end{pmatrix}\begin{pmatrix}
\chi_D \\
\Phi_D\end{pmatrix}\,+\,\mathcal{O}(1/m^2)~. 
\end{align}
The equation in the first row of (\ref{eq:uniteqexp}) 
yields
\begin{align}
   \chi_E\,=\,\chi_D\,-\, i\,\frac{g}{2\,m}\,a\,\Phi_D~, 
\end{align}
where $\Phi_D$ 
can be expressed in terms of $\chi_D$ by using 
the Dirac equation:
\begin{align}
    \Phi_D\,=\,-\hat{D}_D^{-1}[\hat{C}_D[\chi_D]]=\frac{1}{2\,m}\,\vec{\sigma}\cdot\hat{\vec{P}}\,\chi_D+\,\mathcal{O}(1/m^2)~,
\end{align}
with\footnote{Here, we use the $D$ subscript for $\hat{C}$ and $\hat{D}$ to make it clear that they refer to the derivative basis.}
\begin{align}
\hat{C}_D\,&=\,\vec{\sigma}\cdot\hat{\vec{P}}\,-\,\frac{g}{2\,m}\,\dot{a}~,\nonumber\\   
\hat{D}_D\,&=\,-\left(2\,m\,+\,\hat{\epsilon}\,\,+\,\frac{g}{2\,m}\,\vec{\sigma}\cdot\vec{\nabla}a\,+\,e\,A_0\right)~.
\end{align}
This results in 
\begin{align}\label{eq:relates2basis}
   \chi_E\,=\,\chi_D\,-\, i\,\frac{g\,a}{4\,m^2}\,\vec{\sigma}\cdot\hat{\vec{P}}\,\chi_D~.
\end{align}
\eq{eq:relates2basis} can be understood as the 
linear term of an exponentiation, which reconstructs 
the unitary transformation 
\begin{align}\label{eq:relates2basisUnit}
   \chi_E\,=
   \exp\left\lbrace -i \frac{g\,a}{4\,m^2} \vec\sigma \cdot \hat{\vec P}\right\rbrace \,\chi_D~.  
\end{align}
A unitary transformation does not alter the underlying physics, 
as long as both the states and the Hamiltonian 
are transformed.
Applying the transformation in \eq{eq:relates2basisUnit} 
to the Schr\"odinger equation
reveals a shift in the Hamiltonian $\hat{\H}_a^{(2)}$ of the following form: 
\begin{align}
    \hat{\H}^{(2)}_{Ea}[\chi_E]\,&=\,\hat{\H}_a^{(2)}[\chi_E]\,-\,i\,\partial_t\left(\text{exp}\left\lbrace\,+i\,\frac{g\,a}{4\,m^2}\,\vec{\sigma}\cdot\hat{\vec{P}}\right\rbrace\chi_E\right)\nonumber\\
    &\,-\,e\,A_0\,\text{exp}\left\lbrace+\,i\,\frac{g\,a}{4\,m^2}\,\vec{\sigma}\cdot\hat{\vec{P}}\right\rbrace\chi_E + \mathcal{O}(1/m^4) \, , 
\end{align}
which, after expanding to $\mathcal{O}(1/m^2)$ and using 
\begin{align}
    i\,\partial_t\,\chi_E\,=\,-\,e\,A_0\,\chi_E\,+\,\mathcal{O}(1/m)~,
\end{align}
results in
\begin{align}\label{eq:HU2}
    \hat{\H}^{(2)}_{Ea}\,=\,-\,\frac{g\,\lbrace\vec{\sigma}\cdot\hat{\vec{P}},\dot{a}\rbrace}{4\,m^2}\,
    -\,\frac{e\,g}{4\,m^2}
    a\,\vec{\sigma}\cdot\vec{E}
    +\frac{g}{4m^2} \dot{a} \vec{\sigma}\cdot\hat{\vec{P}}~,
\end{align}
which precisely reproduces the exponential basis 
Hamiltonian in \eq{eq:Hexp}. 
This argument can also be seen as a non-trivial check 
of the more involved calculation of the 
axion Hamiltonian in the exponential basis, 
that is provided in \app{app:equivalence}. 

More generally, it is possible to perform 
the following unitary transformation 
on the state in the derivative basis \cite{Smith:2023htu}  
\begin{align}\label{eq:relates2basisUnit2}
   \chi_U\,=
   \exp\left\lbrace - i \beta \frac{g}{4\,m^2} 
   \lbrace\vec{\sigma}\cdot\hat{\vec{P}},a\rbrace
   \right\rbrace \,\chi_D~,   
\end{align}
with $\beta$ being a free parameter. 
Following similar steps as for the derivation below 
\eq{eq:relates2basisUnit}, one arrives at the 
following Hamiltonian 
\begin{align}\label{eq:HU22}
    \hat{\H}^{(2)}_U\,=\,-\,(1-\beta)\,\frac{g\,\lbrace\vec{\sigma}\cdot\hat{\vec{P}},\dot{a}\rbrace}{4\,m^2}\,
    -\,\beta\,\frac{e\,g}{2\,m^2} a\,\vec{\sigma}\cdot\vec{E}~.
\end{align}
In particular, choosing $\beta = 1$, 
the Hamiltonian in \eq{eq:HU22} corresponds 
to that of  
an axion-dependent EDM \cite{Smith:2023htu}, 
to which we will refer to in the following as the 
``EDM picture''. 
However, in the next section 
it will be show 
that,
when doing time-dependent perturbation theory, 
the transformation of the state 
in \eq{eq:relates2basisUnit2} implies 
the presence of an axion boundary term 
which does not allow 
to interpret the net effect 
of \eq{eq:HU22} with $\beta = 1$
solely in terms of an 
axion-dependent EDM.

\section{Axion EDM in time-dependent perturbation theory}
\label{sec:timedepPT}

To describe the time evolution of a physical system 
in NR quantum mechanics, we can employ 
the interaction picture in which the Hamiltonian is 
split into $\hat\H (t) = \hat\H_0 + \hat\V (t)$, 
where $\hat \H_0 = \hat{\vec{p}}^{\ 2}/(2 m)$ 
is the free Hamiltonian in the NR limit and 
$\V (t)$ is a perturbation which depends explicitly on time, 
encoding in particular the axion dependence.  
The state in the interaction picture is  
defined as
$\ket{\chi_I(t)}\,=\,e^{i\hat\H_0 t}\,\ket{\chi (t)}$ 
and its time evolution is governed by 
$\ket{\chi_I(t)}\,=\,\hat{U}(t,t_0)\,\ket{\chi_I (t_0)}$, 
in terms of the time evolution operator 
\begin{align}
\label{eq:Udef}
    \hat{U}(t,t_0)\,=\,T\, \text{exp}\left\lbrace-\,i\,\int_{t_0}^{t}\text{d}t^\prime\,\hat{\V}_{I}(t^\prime)\right\rbrace~,
\end{align}
where $T$ is the time-ordered product.

In the following, we will adopt the temporal 
gauge\footnote{A crucial simplification arising in the temporal gauge consists in the fact that we can 
describe the axion dynamics at $\mathcal{O}(1/m^2)$ 
by doing time-dependent perturbation at the first order, thus neglecting the 
$\mathcal{O}(1/m^2)$
interference  
with the leading order term $\hat\H (t) = 
-eA_0 + \ldots$ arising at the second order 
in time-dependent perturbation theory.}
$A_0 = 0$ and further consider the approximation 
$\vec \nabla a = 0$, which is phenomenologically motivated in the case of NR dark matter axions, 
for which
$\vec \nabla a \simeq m_a \vec v$ and 
$v/c \simeq 10^{-3}$. 
In such a case, focusing on the 
axion-dependent terms, 
$\hat{\V}(t) = \hat{\H}^{(2)}_{a}$,  
and hence 
$\hat{\V}_{I}(t)=
e^{i\hat\H_0 t} \hat{\V}(t) e^{-i\hat\H_0 t}
= \hat{\V}(t)+\mathcal{O}(1/m^3)$ \cite{Smith:2023htu}. 
Hereafter, we will drop for simplicity
the subscript $I$ when referring to the 
interaction picture.

\subsection{Check of equivalence theorem 
and axion shift symmetry} 

In the spirit of the equivalence theorem in 
quantum field theory \cite{Kamefuchi:1961sb,Chisholm:1961tha,Kallosh:1972ap}, which implies the exact matching of 
$S$-matrix elements in different bases 
connected by field redefinitions, 
it is instructive to 
verify that 
\beq 
\label{eq:equivperttheory}
\bra{\psi_D(t)} \hat U_D (t,t_0) \ket{\chi_D(t_0)} = 
\bra{\psi_E(t)} \hat U_E (t,t_0) \ket{\chi_E(t_0)}
\eeq
holds in perturbation theory, 
where we have explicitly labeled 
the states and the time evolution operator
in the derivative ($D$) and exponential ($E$) bases. 

In order to check \eq{eq:equivperttheory} 
at the first order in time-dependent perturbation theory, 
let us consider first the left-hand side (LHS) 
\begin{align} 
\label{eq:equivalenceLHS}
\text{LHS}\, (\ref{eq:equivperttheory}) &= 
\bra{\psi_D(t)} \[ 1 - i \int_{t_0}^{t} \text{d}t'\,\hat{\V}_{D}(t') \] \ket{\chi_D(t_0)} \\
&= \bra{\psi_D(t)} 
\[ 1 + \frac{ig}{2\,m^2} \vec{\sigma} \cdot
\int_{t_0}^{t} \text{d}t'\,   
(\partial_{t'} a) \, (\hat{\vec{p}} + e \vec{A}(t')) \] 
\ket{\chi_D(t_0)} 
\, , \nonumber 
\end{align}
where in the second step we have employed the axion Hamiltonian 
in the derivative basis from \eq{eq:fuA}. 
On the other hand, taking into account the 
shift of the state from \eq{eq:relates2basis}
\beq 
\label{eq:stateshift} 
\ket{\chi_E (t_0)} = \( 1 - i\,\frac{g\,a(t_0)}{4\,m^2}\,\vec{\sigma}\cdot\hat{\vec{P}}(t_0) \) \ket{\chi_D (t_0)} \, ,  
\eeq
as well as the axion Hamiltonian in 
the exponential basis from \eq{eq:Hexp}, 
the right-hand side (RHS) 
of \eq{eq:equivperttheory} 
can be rewritten as
\begin{align} 
\label{eq:equivalenceRHS}
&\text{RHS}\, (\ref{eq:equivperttheory}) = 
\bra{\psi_E(t)} \[ 1 - i \int_{t_0}^{t} \text{d}t'\,\hat{\V}_{E}(t') \] \ket{\chi_E(t_0)} \nonumber \\
&= \bra{\psi_D(t)} 
\( 1 + i\,\frac{g\,a(t)}{4\,m^2}\,\vec{\sigma}\cdot\hat{\vec{P}}(t) \)
\nonumber \\
&\times \[ 1  
+\frac{ig}{4\,m^2} \vec{\sigma} \cdot
\int_{t_0}^{t} \text{d}t'\,   
(\partial_{t'} a) \, (\hat{\vec{p}} + e \vec{A}(t')) 
+\frac{ieg}{4\,m^2} \vec{\sigma} \cdot
\int_{t_0}^{t} \text{d}t'\,   
a(t') \, \vec{E}(t') 
\]
\nonumber \\
&\times \( 1 - i\,\frac{g\,a(t_0)}{4\,m^2}\,\vec{\sigma}\cdot\hat{\vec{P}}(t_0) \) \ket{\chi_D (t_0)} \nonumber \\
&= \bra{\psi_D(t)} 
\( 1 + i\,\frac{g\,a(t)}{4\,m^2}\,\vec{\sigma}\cdot\hat{\vec{P}}(t) \)
\nonumber \\
&\times \[ 1  
+\frac{ig}{2\,m^2} \vec{\sigma} \cdot
\int_{t_0}^{t} \text{d}t'\,   
(\partial_{t'} a) \, (\hat{\vec{p}} + e \vec{A}(t')) 
-\frac{ig}{4\,m^2} \vec{\sigma} \cdot
\( a(t) \hat{\vec{P}}(t) - a(t_0) \hat{\vec{P}}(t_0) \)
\]
\nonumber \\
&\times \( 1 - i\,\frac{g\,a(t_0)}{4\,m^2}\,\vec{\sigma}\cdot\hat{\vec{P}}(t_0) \) \ket{\chi_D (t_0)} = 
\text{LHS}\, (\ref{eq:equivperttheory})
\, ,  
\end{align} 
where in the second to last step we have taken into account the definition of the electric field in the temporal gauge ($\vec{E} = - \partial_t \vec{A}$), integrated by parts, and used that $\hat{\vec{p}}$ is time-independent to add and subtract a total derivative. Note that to prove the equivalence 
of the two bases  
it was crucial to take into account the 
transformation of the state in \eq{eq:stateshift}, 
which effectively leads to a boundary term 
in the time-integrated Hamiltonian, 
which exactly cancels out the boundary term 
arising from the integration by parts.

Similarly, we can verify that the 
axion shift symmetry is properly 
preserved in time-dependent perturbation theory. 
While this is manifest in the derivative basis (cf.~\eq{eq:PQsymmetryDER} and \eq{eq:fuA}), 
being the 
Hamiltonian itself shift-invariant, 
that is not the case for the exponential basis. 
Since in the exponential basis also the fermion 
field transforms under the $\U(1)_{\rm PQ}$ symmetry 
(cf.~\eq{eq:PQsymmetryEXP}), 
we expect that the invariance under the shift symmetry can be recovered in the matrix element 
$\bra{\psi_E(t)} \hat U_E (t,t_0) \ket{\chi_E(t_0)}$ 
after including the shift of the state. 

To verify this last statement, let us first identify how the $\U(1)_{\rm PQ}$ symmetry acts on the 
state $\ket{\chi_E(t_0)}$. Combining  \eq{eq:PQsymmetryEXP}, (\ref{eq:EtoD}) 
and (\ref{eq:relates2basis}) we have 
\beq 
\label{eq:PQsymmetryEXP2}
\U(1)^{E}_{\rm PQ}: \qquad
a \to a + \alpha \, m / g \, , \quad 
\ket{\chi_E(t_0)} \to \( 1 
+ i\,\frac{\alpha}{4\,m}\,\vec{\sigma}\cdot\hat{\vec{P}}(t_0)\,\) \ket{\chi_E(t_0)} \, . 
\eeq
Hence, applying the shift in \eq{eq:PQsymmetryEXP2} 
to the matrix element in the exponential basis 
at the first order in time-dependent perturbation theory, we find (see also \eq{eq:equivalenceRHS})
\begin{align}
\label{eq:checkshift} 
&\bra{\psi_E(t)} \[ 1 - i \int_{t_0}^{t} \text{d}t'\,\hat{\V}_{E}(t') \] \ket{\chi_E(t_0)} \nonumber \\ 
&\to \bra{\psi_E(t)} 
\( 1 
- i\,\frac{\alpha}{4\,m}\,\vec{\sigma}\cdot\hat{\vec{P}}(t)\,\) \[ 1 
+
i\,\frac{\alpha\, e}{4\,m}
\vec{\sigma} \cdot
\int_{t_0}^{t} \text{d}t'\, 
\big(-
\partial_{t'} \vec{A}
\, \big)
\] 
\( 1 
+ i\,\frac{\alpha}{4\,m}\,\vec{\sigma}\cdot\hat{\vec{P}}(t_0)\,\)
\ket{\chi_E(t_0)} \nonumber \\
&= \bra{\psi_E(t)} \[ 1 - i \int_{t_0}^{t} \text{d}t'\,\hat{\V}_{E}(t') \] \ket{\chi_E(t_0)}
\, ,
\end{align} 
up to $\mathcal{O}(1/m^2)$ corrections. 
Also in this case the transformation 
of the state, 
resulting in an effective boundary term, 
was crucial 
for ensuring invariance 
under the axion
shift symmetry.

\subsection{Physical relevance of the axion boundary term}

We next discuss the physical effects stemming from the 
axion Hamiltonian at $\mathcal{O}(1/m^2)$. 
Let us consider the derivative basis,
and employ as before the temporal gauge $A_0 = 0$ and work in 
the $\nabla a = 0$ approximation. 
Then at the first order in time-dependent perturbation theory one has 
\beq 
\label{eq:Utt0pert}
\hat U (t,t_0) \simeq 1 - i \int_{t_0}^{t} \text{d}t'\,\hat{\V}_{D}(t') 
= 1 + i \frac{g}{2m^2} \int_{t_0}^{t} \text{d}t'\, 
\dot a \, \vec \sigma \cdot (\hat{\vec p} + e \vec A)
\, .  
\eeq
Upon integration by parts, 
the interacting term 
of $\hat U (t,t_0)$ 
in \eq{eq:Utt0pert}
can be rewritten as 
\beq 
\label{eq:intbyparts}
 i\frac{g}{2m^2} \int_{t_0}^t dt' \[ \dot a \,  \vec \sigma \cdot (\hat{\vec p} + e \vec A) \] = 
 i\frac{g}{2m^2} \int_{t_0}^t dt' a \, \vec \sigma \cdot  e\vec E 
 +i\frac{g}{2m^2} \[ a \vec \sigma \cdot (\hat{\vec p} + e \vec A) \]_{t_0}^t \, ,  
\eeq
which leads to the emergence of a standard 
axion EDM 
\cite{Smith:2023htu} as well as an axion boundary term. However, differently from Ref.~\cite{Smith:2023htu}, we here argue that 
the axion boundary term is 
crucial for correctly describing the physical 
effect. 

Note that the LHS of \eq{eq:intbyparts} 
manifestly 
satisfies two important properties: 
$i)$ it preserves the axion 
shift symmetry $a \to a + \alpha \, m/g$ 
and $ii)$ it yields no static EDM.\footnote{The 
absence of a static EDM, e.g.~arising 
from an axion vacuum 
expectation value, $\theta_{\rm eff} = \vev{a}/f_a$, can be also understood by performing 
in the original Lagrangian 
of \eq{eq:Laxionnonder}
a chiral transformation on the fermion field, 
in such a way that 
$\theta_{\rm eff}$ is rotated away 
and no other physical effect is generated.
This is because, being $\theta_{\rm eff}$ constant, 
no extra term is generated by the shift of the 
fermion kinetic term, while the anomalous 
contribution proportional to 
$\theta_{\rm eff} F \tilde F$
is a total derivative which bears no physical 
consequences for an abelian gauge theory.} 
For these two properties to keep holding in the RHS of \eq{eq:intbyparts} it is crucial to keep the 
axion boundary term. 
Let us show 
in turn these two properties
from the standpoint of the RHS of \eq{eq:intbyparts}: 
\begin{itemize}
\item Shift invariance ($a \to a + \alpha \, m/g$)
\beq 
\text{RHS}\,(\ref{eq:intbyparts}) \to \text{RHS}\,(\ref{eq:intbyparts}) + 
\underbrace{i \frac{\alpha}{2m} \[ \int_{t_0}^t dt' \vec \sigma \cdot  (- e\partial_{t'} \vec A)  + \[ \vec \sigma \cdot (\hat{\vec p} + e \vec A) \]_{t_0}^t \] }_{0}  \, .
\eeq
\item No static EDM ($a = a_0$)
\beq 
\label{eq:no static EDM}
\text{RHS}\,(\ref{eq:intbyparts}) = i\frac{g a_0}{2m^2} \[  \int_{t_0}^t dt' \vec \sigma \cdot ( - e\partial_{t'} \vec A)  + \[ \vec \sigma \cdot (\hat{\vec p} + e \vec A) \]_{t_0}^t   \] = 0 \, .
\eeq
\end{itemize}
Hence, this argument suggests that the axion 
boundary term should also play a crucial role in the phenomenology of the oscillating axion EDM. 

Finally, it is worth noting that the results 
presented in this section can be 
equivalently  
obtained 
in the ``EDM picture'' discussed at the end 
of \sect{sec:unitary},  
upon 
taking into account the shift of the external states
arising from \eq{eq:relates2basisUnit2} with $\beta = 1$. 
This effectively corresponds to the inclusion of an axion boundary term, which exactly matches the RHS of \eq{eq:intbyparts}.

\subsection{Effective axion EDM}

To assess the physical consequences of the axion 
boundary term 
of \eq{eq:intbyparts}
for the EDM phenomenology, 
let us assume the following configuration
featuring a constant electric 
field
as well as an oscillating axion dark matter field
\beq 
\vec E = \text{const} \, , \qquad 
\vec A = - t \vec E \, , \qquad  
a(t) = a_0 \cos(m_a t \, + \, \theta_0) \, ,  \label{eq:conditions}
\eeq
where we set $t_0 = 0$, so that $t$ is the measurement time. Here, $a_0$ and $\theta_0$ are the amplitude and initial phase of the axion field, respectively. Below, without loss of generality, we set $\theta_0=0$.
The amplitude of axion oscillations is
fixed in terms of the local dark matter relic density as 
\beq 
\label{eq:axionDMampl}
a_0 = \frac{\sqrt{2\,\rho_{\rm DM}}}{m_a} \, , 
\eeq
where $\rho_{\rm DM} \simeq 0.3 \, \text{GeV}/\text{cm}^3 = 2 \cdot 10^{-42}\, \text{GeV}^4$~\cite{ParticleDataGroup:2022pth}. 
For a QCD axion, this implies $a_0/f_a \simeq 4 \times 10^{-17}$ irrespective of the axion mass and decay constant.

We are working in a perturbative expansion about a free state, which together with the non-relativistic expansion implies the range of validity $|e \vec{E} t| \ll |\vec{p}_0| \ll m $, where $\vec{p}_0$ is the initial momentum of the fermion. This implies that our perturbative calculation breaks down once $\vec{E}$ has had sufficient time to significantly alter the fermion momentum. With this setup in mind, \eq{eq:Utt0pert} becomes
\begin{align}
\hat U (t,0) \simeq 1
+ i\frac{g a_0}{2m^2} \vec\sigma \cdot \[ \frac{e \vec E}{m_a} \Big( \sin (m_a t) - m_a t \cos (m_a t) \Big) + \hat{\vec p} \( \cos (m_a t) - 1\) \] \, . 
\end{align}
In this expression, we are particularly interested in the 
term proportional to $\vec{\sigma} \cdot \vec{E}$, which can be matched onto an EDM. To this end, it is useful to rewrite such a term in the form of an effective 
time-integrated EDM, defined via the following expression
\begin{align}
\label{eq:RHSEconst}
\hat U (t,0) &\supset 1
+ i\frac{e g a_0}{2m^2}\frac{1}{m_a} \Big( \sin (m_a t) -  \,m_a t \cos (m_a t) \Big) \vec\sigma \cdot \vec E \nonumber \\
&\equiv 1 + i \[ \int_0^t d_{\rm eff} (a (t')) \, dt' \]  \vec\sigma \cdot \vec E\, .
\end{align}
To deal with a quantity with 
the dimension 
of an EDM, we introduce the time average 
\beq 
\label{eq:timeaveragedEDM} 
\vev{d_{\rm eff}} \equiv \frac{1}{t} 
\[ \int_0^t d_{\rm eff} (a (t')) \, dt' \] 
= \frac{e g a_0}{2m^2}\frac{1}{m_a t} \Big( \sin (m_a t) -  \,m_a t \cos (m_a t) \Big) \, , 
\eeq
where the last step follows 
directly
from 
\eq{eq:RHSEconst}. 
Note that the first term 
in the RHS of \eq{eq:timeaveragedEDM}
corresponds to 
the standard oscillating EDM 
(discussed e.g.~in \cite{Smith:2023htu}), 
while the second term 
can be understood to 
arise from the 
axion boundary term.

Let us consider now the following three oscillation regimes: 
\begin{enumerate}
\item Slow oscillations ($m_a t \ll 1$).

The time-averaged effective EDM
can be approximated as
\begin{align}
\label{eq:effectiveEDM}
\vev{d_{\rm eff}}  
&\simeq  
\frac{e g a_0}{2m^2}  
\frac{1}{3} (m_a t)^2
\, ,
\end{align}
and in the case of no oscillations, $m_a t \to 0$, the static EDM contribution goes to zero. 
Note that in the $m_a t \ll 1$ regime, 
the axion boundary term cancels the leading 
order contribution to the standard EDM 
and the overall contribution is suppressed by $(m_a t)^2 \ll 1$. So we conclude that it is crucial to include the axion boundary term to correctly interpret the physical effect of the 
time-averaged effective EDM. 
For non-zero initial axion phase $\theta_0$, $\vev{d_{\rm eff}}$ is linear in $m_a t$ and still goes to zero.

\item Intermediate oscillations ($m_a t \sim 1$). 

In this regime, in \eq{eq:timeaveragedEDM} both the standard EDM contribution 
and the one arising from the axion boundary term 
are comparable, corresponding to an effective EDM amplitude  
\beq 
\label{eq:EDMstrength}
\vev{d_{\rm eff}} \sim \frac{e g a_0}{2m^2} = \frac{e}{2m} 
\(\frac{a_0}{f_a}\) \, .  
\eeq
In the electron case, 
taking a unit axion-electron coupling, we obtain 
$\vev{d_{\rm eff}} \sim 7 \cdot 10^{-30} \ \text{e cm}$.
This is notably
of the same order as the static electron EDM bound,  
$|d_e| \lesssim 1.1 \cdot 10^{-29}$ e cm 
\cite{ACME:2018yjb}.

\item Fast oscillations ($m_a t \gg 1$).

In this case the boundary term 
contribution
in \eq{eq:effectiveEDM} always dominates, thus leading 
to a  
time-averaged effective 
EDM  
that avoids the $1/(m_a t) \ll 1$ suppression of
the standard axion EDM contribution and maintains the amplitude of \eq{eq:EDMstrength}:
\begin{align}
\label{eq:RHSEconsfastt}
\vev{d_{\rm eff}}  \simeq 
- \frac{e g a_0}{2m^2} \cos (m_a t)
\, . 
\end{align}
\end{enumerate}
\begin{figure}[t!]
    \centering    \includegraphics{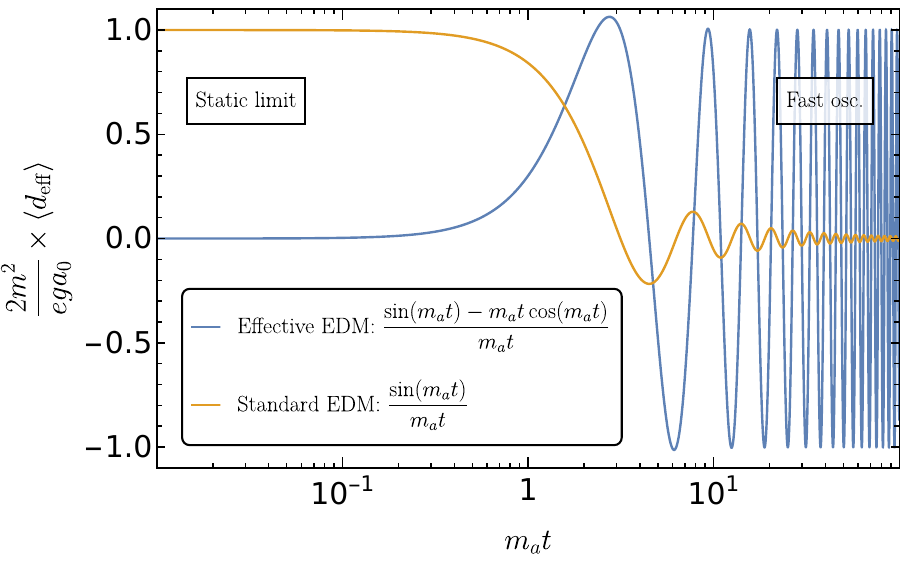}
    \caption{\label{fig:effectiveEDM} 
    Time-averaged effective EDM including 
    the axion boundary term (blue) vs.~standard EDM (yellow), 
    for a constant electric field 
    as given by \eq{eq:timeaveragedEDM}. 
    In the static limit, $m_a t \to 0$, the effective EDM vanishes when taking the axion boundary term into account. 
    In the rapidly oscillating regime, $m_a t \gg 1$, the axion boundary term dominates and makes the amplitude 
    of the effective EDM independent of $m_a t$, 
    while in contrast the contribution of the standard EDM decouples as $1/(m_a t)$.
    }
\end{figure}
The behavior of the 
time-averaged effective
EDM  
of \eq{eq:timeaveragedEDM}
is displayed in \fig{fig:effectiveEDM}  
in comparison to the 
standard EDM contribution 
discussed e.g.~in Ref.~\cite{Smith:2023htu}, i.e.~without axion 
boundary term. 
We have verified that, as expected from \eq{eq:no static EDM}, the effective EDM decouples irrespective of the initial axion phase $\theta_0$ because of a cancellation with the boundary term.
In the absence of this boundary term, the EDM does not vanish in the static limit but rather takes a generally non-zero value depending on $\theta_0$. 

When interpreting these results, 
it is important to keep in mind the range of validity of the perturbative expansion. Ultimately, the range of validity of our calculation is limited to, taking again the case of an electron,
\begin{gather}
    m_a t \ll \frac{m_a m}{e |\vec{E}|} \simeq 
    2.6
    \( \frac{10^6 \, V/\text{m}}{|\vec{E}|} \) 
    \( \frac{m_a}{\mu\text{eV}} \)
    \, .
\end{gather}
We conclude that the 
axion boundary term contribution 
crucially affects the phenomenology 
of the oscillating EDM 
by suppressing the effect in the small oscillation regime and by enhancing it in the regime of fast oscillations. 
Remarkably, for an order one axion-electron coupling, as predicted in benchmark axion models \cite{Zhitnitsky:1980tq,Dine:1981rt}, and oscillations either intermediate or fast, the amplitude of the effective oscillating EDM is of the order of the present static EDM limit.
However, a proper discussion of the experimental prospects 
for measuring such effect is beyond 
the scope of the present paper and it is left for future work.  

\section{Conclusions}
\label{sec:concl}

In this paper, we re-examined the 
recent claim \cite{Smith:2023htu} 
(see also \cite{Alexander:2017zoh,Wang:2021dfj})
that the 
axion-fermion coupling is responsible for 
an oscillating EDM in the background of axion dark matter, cf.~\eq{eq:aedm}. 
By employing the Pauli elimination method for the NR expansion of the axion Dirac equation, we provided an alternative derivation of the axion Hamiltonian, 
emphasizing the equivalence between the derivative and exponential bases. 

Unlike previous studies, we pointed out the 
physical relevance of an axion boundary term, which 
turns out to be crucial in order to restore the axion shift symmetry and plays a critical role for the 
oscillating EDM phenomenology, e.g.~by 
leading to an exact cancellation of the 
standard EDM contribution 
in the static limit. 

In the case of a constant electric field, 
we were able to introduce the notion of 
a time-averaged effective
EDM, which comprises 
both the standard contribution of \eq{eq:aedm} 
and the axion boundary term. 
As exemplified by \fig{fig:effectiveEDM}, 
different patterns emerge
depending on the 
oscillation regime.  
For slow oscillations,  
the effective EDM is suppressed compared to the standard one, while for fast 
oscillations it is 
relatively enhanced, avoiding the suppression experienced by the standard EDM contribution.
Only in the intermediate oscillation regime 
are the effective and standard EDM 
contributions comparable. 

This new observable 
is particularly relevant for 
the case of the electron EDM, 
since for an $\mathcal{O}(1)$ axion-electron coupling the amplitude of the effective EDM is comparable to the present static EDM limit if in  a regime of intermediate or fast oscillations. 
The experimental verification of such a scenario remains an interesting open question, which will be addressed elsewhere.

\section*{Note added} 

Simultaneously with our initial submission 
to the arXiv, 
a revised version of Ref.~\cite{Smith:2023htu} 
appeared, reaching the same conclusions regarding the decoupling of the EDM in the static limit.  

\section*{Acknowledgments}
 
We thank Giovanni Carugno for bringing 
this problem to 
our attention,   
as well as  
Ramona Gr\"ober, 
Sebastian Hoof, 
Gino Isidori, 
Marco Peloso,  
Pierre Sikivie 
and 
Christopher Smith 
for useful discussions. 
This work is funded by the European Union -- NextGenerationEU and by the University of Padua under the 2021 STARS Grants@Unipd programme (Acronym and title of the project: CPV-Axion -- Discovering the CP-violating axion). 
The work of LDL is also supported by the European Union -- Next Generation EU and by the Italian Ministry of University and 
Research (MUR) via the PRIN 2022 project n.~2022K4B58X -- AxionOrigins. 

\appendix 

\section{Axion Hamiltonian in the exponential basis}
\label{app:equivalence}

Let us consider the exponential basis of \eq{eq:Laxionnonder},  
in which the axion Dirac equation reads 
\beq 
\label{eq:Diraceqnonlin}
\left(i \gamma^\mu\partial_\mu - m\,\cos\left(\frac{g\,a}{m}\right)\, + e\gamma^\mu A_\mu - i m\,\sin\left(\frac{g\,a}{m}\right)\, \gamma_5\right)\,\psi = 0 \, .
\eeq 
In terms of two-component spinors, as in \eq{eq:Diracmatrix}, 
the axion Dirac equation yields
\begin{align}\label{eq:ABCDexp}
\begin{aligned}
    \hat{A}&=\hat\epsilon+m-m\cos(\frac{g}{m}a) + e A_0~,\\
    \hat{B}\,&=\,-\vec{\sigma}\cdot\hat{\vec{P}} - i \,m\,\sin\left(\frac{g\,a}{m}\right)~,\\
    \hat{C}\,&=\,\vec{\sigma}\cdot\hat{\vec{P}} - i \,m\,\sin\left(\frac{g\,a}{m}\right)~,\\
    \hat{D}&=-\left(\hat\epsilon +m+m\cos(\frac{g}{m}a) + e A_0\right)~.
\end{aligned}
\end{align}
Note that here $\psi=\exp(-i\ m\ t)\ (\chi_{\slashed{N}},\ \Phi)^T$ and $\hat{A},\hat{B},\hat{C}$ and $\hat{D}$ are distinct from their analogs in the derivative basis. Nevertheless, in the interest of simplicity, we will neglect distinguishing subscripts in this appendix. The meaning of the subscript $_{\slashed{N}}$ is related to normalization and will become clear later. As in the derivative basis, the Pauli elimination method leads to solutions in the form of \eq{eq:DiracAxion}. Expanding the inverse of $\hat{D}$ in powers of $1/m$, we obtain the same result as \eq{eq:Dinverse} at $\mathcal{O}(1/m^2)$. 
As in the derivative basis, we match onto the form of a Schr\"odinger equation, see \eq{eq: Schrodinger}, obtaining the same axion Hamiltonian at $\mathcal{O}(1/m)$ given by \eq{eq:nonormalizedaxionEXPbasis}. 

It is worth mentioning that by working in the linear 
axion basis instead, i.e.~approximating 
$m\cos(ga/m) \simeq ga$ and $m\sin(ga/m) \simeq ga$ in \eq{eq:Diraceqnonlin}, one would have obtained 
\begin{align}\label{eq:Hlinbasis}
    \hat{\H}^{(1),\, \rm linear}_{a}\,=
    \,-\frac{g}{2\,m}\vec{\sigma}\cdot\vec{\nabla}a + \frac{g^2 a^2}{2m}~, 
\end{align}
featuring a spurious term that breaks the axion shift symmetry. 
Thus, we stress the importance of properly taking into account the 
non-linear axion 
terms arising from the exponential axion parametrization at a given order in the NR 
expansion.\footnote{This is in contrast e.g.~with Refs.~\cite{Alexander:2017zoh,Wang:2021dfj},  
whose results are obtained in the linear basis.} 

The previous calculation in the exponential basis can be 
extended at $\mathcal{O}(1/m^2)$. 
From Eq.~\eqref{eq:DiracAxion} we obtain:
\begin{align}
\label{eq:expansatz}
&i\,\partial_t\,\chi_{\slashed{N}}\,+\,e\,A_0\,\chi_{\slashed{N}}\,-\,\hat{\H}^{(1)}[\chi_{\slashed{N}}]\,-\,f^{(2)}(\chi_{\slashed{N}},\partial_t\,\chi_{\slashed{N}})\,+\,\mathcal{O}(1/m^3)\,=\,0~, 
\end{align}
where $f^{(2)}(\chi_{\slashed{N}},\partial_t\,\chi_{\slashed{N}})$ is an $\mathcal{O}(1/m^2)$ function that depends explicitly on $\chi_{\slashed{N}}$ and $\partial_t\,\chi_{\slashed{N}}$. Note that due to the dependence on $\partial_t\,\chi_{\slashed{N}}$ at $\mathcal{O}(1/m^2)$, the exponential basis presents a distinct challenge, as it does not allow \eq{eq:expansatz} to be expressed as a Schr\"odinger-like equation. 
To eliminate 
the contribution of $\partial_t\,\chi_{\slashed{N}}$ from this expression 
we can proceed iteratively, by employing the initial portion of 
\eq{eq:expansatz},
\begin{align}
    i\,\partial_t\,\chi_{\slashed{N}}\,=\,-\,e\,A_0\,\chi_{\slashed{N}}\,+\,\mathcal{O}(1/m)~.\label{eq:leading order hamiltonian}
\end{align}
Note that there is no need to include here corrections of order $\mathcal{O}(1/m)$ since they will only contribute to the Hamiltonian at order $\mathcal{O}(1/m^3)$. By performing this iteration, we arrive at the following expression for the Hamiltonian at $\mathcal{O}(1/m^2)$ in the exponential basis:
\begin{align}
\hat{\H}^{(2)}_{a\slashed{N}}\,=\,&-\, \frac{e \,g }{4\, m^2}\,a\,\vec{\sigma}\cdot\vec{E}\,-\, \frac{
 i\, g^2}{4 \,m^2}\,a\, \dot{a}\,-\, \frac{e\, g}{4\, m^2}\,\dot{a}\,\vec{\sigma}\cdot\vec{A} -\,\frac{ g}{4\, m^2}\,\, \dot{a}\, \vec{\sigma} \cdot \hat{\vec{p}} +\, \frac{i\,
 g}{4\, m^2}\,\vec{\sigma}\cdot\vec{\nabla}\dot{a}~. \label{eq:hamiltonian,exponential basis,not normalized}
\end{align}
In the first term of \eq{eq:hamiltonian,exponential basis,not normalized}, 
we see the emergence of an axion-dependent EDM. 
However, this Hamiltonian features the term $ig^2a\dot{a}/(4m^2)$ which is not Hermitian. This means that if this were to be the 
actual Hamiltonian, then the normalization 
of the state
would not be preserved in time.\footnote{Since the problem of non-Hermiticity appears at $\mathcal{O}(1/m^2)$, we expect a normalization operator that enters at exactly this order and that the Hamiltonian corresponding to the correctly normalized spinor will be Hermitian. A normalization that enters only at $\mathcal{O}(1/m^2)$ is further motivated by the fact that the exponential basis already agrees with the derivative basis at $\mathcal{O}(1/m)$.\label{foot:App1}} This is indeed the meaning of $\chi_{\slashed{N}}$ and $\hat{\H}^{(2)}_{a\slashed{N}}$, whose subscript of which indicates that the field is not correctly normalized. 

To address this, consider that  
the hermiticity of the Dirac Hamiltonian implies 
that the norm of the state is constant in time, 
which in coordinate space reads 
\beq 
\label{eq:Diracnormaliz}
\int\dd[3]{x}  \psi^\dag \psi = \text{const} \, .
\eeq
This corresponds to the conservation of the 
global Noether current, associated with the fermion number, 
i.e.~$\partial_\mu (\bar \psi \gamma^\mu \psi) = 0$, from which 
one obtains the continuity equation 
$i \partial_t (\psi^\dag \psi) = - \vec \nabla \cdot (\bar\psi \vec \gamma \psi)$. Upon integrating the latter equation
in $d^3x$ one readily obtains \eq{eq:Diracnormaliz}, 
which can be also rewritten in terms of the large and small component spinors, i.e. 
\beq 
\label{eq:Diracnormalizlargesmall}
\int\dd[3]{x}  \( \chi_{\slashed{N}}^\dag \chi_{\slashed{N}} + \Phi^\dag \Phi \) \, = \text{const} \, , 
\eeq
and substituting $\Phi = - \hat{D}^{-1}[\hat{C}[\chi_{\slashed{N}}]]$ in the 
previous equation we obtain 
\begin{gather}
	\int\dd[3]{x}  \psi^\dag \psi = \int\dd[3]{x} \left[\chi_{\slashed{N}}^\dag\chi_{\slashed{N}} + \left(\hat{D}^{-1}[\hat{C}[\chi_{\slashed{N}}]]\right)^\dagger\left(\hat{D}^{-1}[\hat{C}[\chi_{\slashed{N}}]\right)\right].\label{eq:ub written out}
\end{gather}
If we were able to reduce \eq{eq:ub written out} to the form
\begin{gather}
\int\dd[3]{x}  (\hat{N}[\chi_{\slashed{N}}])^\dag (\hat{N}[\chi_{\slashed{N}}])=\int\dd[3]{x}  \chi^\dag \chi = \text{const.},
\end{gather}
then we would have shown that the normalization of the field $ \chi =\hat{N}[\chi_{\slashed{N}}] $ is correctly preserved in time.
This is non-trivial to achieve because of the derivative (and thus directional) nature of the operators $ \hat{D}^{-1}\hat{C} $. Therefore, we will first evaluate eq.~\eqref{eq:ub written out} in terms of eigenvalues and then seek an operator $ \hat{N} $ that produces the same eigenvalues.
We thus evaluate $ \hat{D}^{-1}[\hat{C}[\chi_{\slashed{N}}]] = \xi \chi_{\slashed{N}} $ to find the eigenvalue $ \xi $. Explicitly, this is
\begin{gather}
	\hat{D}^{-1}[\hat{C}[ \chi_{\slashed{N}}]] = \xi \chi_{\slashed{N}} = \left( \frac{i a g}{2 m}-\frac{e \vec{\sigma}\cdot \vec{A} }{2 m}-\frac{\vec{\sigma}\cdot \vec{p} }{2 m} + \frac{\dot{a} g}{4 m^2}
 -\frac{i e \vec{\sigma}\cdot \vec{E} }{4 m^2}\right)\chi_{\slashed{N}}+\mathcal{O}\left(\frac{1}{m^3}\right),
\end{gather} 
such that $ \xi $ no longer contains any derivative acting outside the operator. Here we made the replacement,
\begin{gather}
	i\vec{\nabla} \chi_{\slashed{N}} = - \vec{p} \chi_{\slashed{N}},
\end{gather}
and applied \eq{eq:leading order hamiltonian}. In this case, it does not matter in which basis we calculate $ \hat{\H} $ because both the derivative and exponential bases agree to $ \mathcal{O}(1/m) $. For this step to be consistent, we have to verify a posteriori that also the Hamiltonian for the normalized spinor agrees up to  $ \mathcal{O}(1/m) $, which will turn out to be the case.
In terms of this eigenvalue $ \xi $, we can write
\begin{align}
	\int\dd[3]{x}  \tilde{\psi}^\dag \tilde\psi &= \int\dd[3]{x} \left[\chi_{\slashed{N}}^\dag\chi_{\slashed{N}} + \left(\xi\chi_{\slashed{N}}\right)^\dagger\left(\xi\chi_{\slashed{N}}\right)\right] \nonumber \\
	&=\int\dd[3]{x} \left[\chi_{\slashed{N}}^\dag\chi_{\slashed{N}} + \chi_{\slashed{N}}^\dag\xi^\dagger\xi\chi_{\slashed{N}}\right] \nonumber \\
	&=\int\dd[3]{x} \left[\left(N\chi_{\slashed{N}}\right)^\dagger\left(N\chi_{\slashed{N}}\right)+\mathcal{O}\left(\frac{1}{m^3}\right) \right],
\end{align}
where we defined the eigenvalue
\begin{gather}
	N\equiv 1+\frac{1}{2}\xi^\dagger\xi = 1+\frac{1}{8m^2}\left(a^2g^2+e^2 \vec{A}^2+\vec{p}^2+e \left\{\vec{\sigma}\cdot\vec{p},\vec{\sigma}\cdot\vec{A}\right\}\right) \, .
\end{gather}
Before promoting $ N $ to a proper operator we have to address the last term, because simply promoting $ \vec{p}\to\hat{\vec{p}} = -i \vec{\nabla} $ would introduce unintended terms from the resulting product rule. To address this, we can apply the anticommutation rule $ \left\{\vec{\sigma}\cdot\vec{x},\vec{\sigma}\cdot\vec{y}\right\} = 2 \vec{x}\cdot\vec{y} $, such that 
\begin{gather}
N\equiv  1+\frac{1}{8m^2}\left(a^2g^2+e^2 \vec{A}^2+\vec{p}^2+2e \vec{A}\cdot\vec{p}\right).
\end{gather}
This is straightforward to promote to an operator form, 
\begin{gather}
\hat{N} = 1+\frac{1}{8m^2}\left(a^2g^2+e^2 \vec{A}^2+\vec{\nabla}^2-2ie \vec{A}\cdot\vec{\nabla}\right),
\end{gather}
which correctly returns the eigenvalue $ N $ when applied to $ \chi_{\slashed{N}} $. 
By inverting the normalization operator, we finally obtain
\begin{gather}
\hat{N}^{-1} = 1-\frac{1}{8m^2}\left(a^2g^2+e^2 \vec{A}^2+\vec{\nabla}^2-2ie \vec{A}\cdot\vec{\nabla}\right),\label{eq:Nin}
\end{gather}
such that $ \chi_{\slashed{N}} = \hat{N}^{-1}[\chi]$. Since this operator depends on $a$ it is clear that it introduces corrections also to the axion-dependent part of the Hamiltonian in the exponential basis.

Let us now see if any contributions were missed in the derivative basis. One can proceed similarly and we find the following:
\begin{align}
 \text{Derivative basis:}\quad   \hat{N}^{-1}\,=\,1-\frac{1}{8m^2}\left(e^2 \vec{A}^2+\vec{\nabla}^2-2ie \vec{A}\cdot\vec{\nabla}\right)\ ,
\end{align}
where, in contrast to the exponential basis, there is no axion dependence at  $\mathcal{O}(1/m^2)$. This can be clearly seen by inspecting Eq.~\eqref{eq:ABCD} where the axion contributions enter at $\mathcal{O}(1/m)$ in contrast to Eq.~\eqref{eq:ABCDexp} where the axion contribution already enters at $\mathcal{O}(1)$. As speculated earlier, this means that the axion-dependent part of the Hamiltonian will not receive any corrections from normalization at $\mathcal{O}(1/m^2)$ in the derivative basis.

Now that we have identified a spinor
that is appropriately normalized, $\chi=\hat{N}[\chi_{\slashed{N}}]$, we want to derive how the corresponding Hamiltonian is modified. To this end, we rewrite $\hat{\H}_{\slashed{N}}[\chi_{\slashed{N}}] = i\partial_t[\chi_{\slashed{N}}]$ as 
\begin{gather}
\hat{\H}_{\slashed{N}}[\hat{N}^{-1}[\chi]] = i\partial_t[ \hat{N}^{-1}[\chi]]~,\label{eq:rewritten Hamiltonian, not expanded}
\end{gather}
and bring this into the canonical form $\hat{H} \chi = i \partial_t \chi$. To this end, let us consider the expansions 
\begin{align}
    \hat{\H}_{\slashed{N}}&= \hat{H}_{\slashed{N}}^{(0)}+ \frac{1}{m}\hat{H}_{\slashed{N}}^{(1)}+ \frac{1}{m^2}\hat{H}_{\slashed{N}}^{(2)} +\mathcal{O}(1/m^3), \\
    \hat{N}^{-1} &= \hat{\alpha}_0+ \frac{1}{m}\hat{\alpha}_1+ \frac{1}{m^2}\hat{\alpha}_2 +\mathcal{O}(1/m^3).
\end{align}
From \eq{eq:Nin}, we see that $\hat{\alpha}_0=1$ and $\hat{\alpha}_1=0$ , which verifies our earlier expectations (cf.~discussion 
in footnote \ref{foot:App1}). Furthermore, we know that normalization only changes terms at $\mathcal{O}(1/m^2)$, so $\hat{H}_{\slashed{N}}^{(0)}=\hat{H}^{(0)}$ and $\hat{H}_{\slashed{N}}^{(1)}=\hat{H}^{(1)}$.
In terms of these parameters, \eq{eq:rewritten Hamiltonian, not expanded} becomes 
\begin{align}\label{eq:normalizationexplore}
i\,\partial_t[\chi] = \, &\hat{H}^{(0)}[\chi] - \frac{1}{m}\,\hat{H}^{(1)}[\chi] \nonumber\\
&- \frac{1}{m^2}\,\left( i\, \partial_t[\hat{\alpha}_2[\chi]] - \hat{H}^{(0)}[\hat{\alpha}_2[\chi]]  - \hat{H}_{\slashed{N}}^{(2)}[\chi] \right) + \mathcal{O}(1/m^3) = 0~. 
\end{align}
In \eq{eq:normalizationexplore} $\hat{\alpha}_2$ only appears with $\hat{H}^{(0)}$ which does not contain the axion. Therefore, only the $-a^2g^2/8 $ term from \eq{eq:Nin} will contribute to the axion-dependent terms at $\mathcal{O}(1/m^2)$.
We then find that
\begin{align}
\hat{\H}_{a}^{(2)}[\chi]&=\frac{1}{8m^2}\left(i\, \partial_t[a^2g^2\chi] - H^{(0)}[a^2g^2\chi] \right)+\hat{\H}_{a\slashed{N}}^{(2)}[\chi] =\frac{ig^2}{4m^2}a\dot{a}\chi + \hat{\H}_{a\slashed{N}}^{(2)}[\chi] \, ,\label{eq:Ha2 in exponential basis, appendix}
\end{align}
which exactly cancels the non-Hermitian term in \eq{eq:hamiltonian,exponential basis,not normalized} 
and yields the axion Hamiltonian in the exponential basis 
as defined in \eq{eq:Hexp}. Note that when we in the main text of this work refer to fields and Hamiltonians defined in the exponential basis we will add the subscript $_E$ to the results above.

\begin{small}

\bibliographystyle{utphys}
\bibliography{bibliography.bib}

\end{small}

\end{document}